\documentclass[12pt,a4paper,final]{iopart}
\usepackage{iopams}
 \expandafter\let\csname equation*\endcsname\relax 
 \expandafter\let\csname endequation*\endcsname\relax 
\usepackage{amsmath,amssymb}  
\usepackage{graphicx}
\usepackage[breaklinks=true,colorlinks=true,linkcolor=blue,urlcolor=blue,citecolor=blue]{hyperref}

\begin{document}

\title[General finite-size effects in
integrable models]{General finite-size effects for zero-entropy
  states in one-dimensional quantum integrable models}

\author{Sebas Eli\"ens}
\address{Institute for Theoretical Physics, Institute of Physics,
  University of Amsterdam, The Netherlands}
 
\ead{I.S.Eliens@UvA.nl}

\author{Jean-S\'ebastien Caux}
\address{Institute for Theoretical Physics, Institute of Physics,
  University of Amsterdam, The Netherlands}
 
\ead{J.S.Caux@UvA.nl}

\begin{abstract}
  We present a general derivation of the spectrum of excitations for
  gapless states of zero entropy density in Bethe ansatz solvable
  models. Our formalism is valid for an arbitrary choice of bare
  energy function which is relevant to situations
  where the Hamiltonian for time evolution differs from the
  Hamiltonian in a (generalized) Gibbs ensemble, i.e. out of
  equilibrium. The energy of particle and hole excitations, as
  measured with the time-evolution Hamiltonian, is shown to
  include additional contributions stemming from the shifts of the
  Fermi points that may now have finite energy. The finite-size
  effects are also derived and the connection with conformal field
  theory discussed. The critical exponents can still be obtained from
  the finite-size spectrum, however the velocity occurring here differs
  from the one in the constant Casimir term. The derivation highlights the importance of the phase
  shifts at the Fermi points for the critical exponents of asymptotes
  of correlations. We generalize certain results known for the ground
  state and discuss the relation to the dressed charge (matrix).
  Finally, we discuss the finite-size corrections in the presence of
  an additional particle or hole which are important for dynamical
  correlation functions.
%We also discuss the case where a single particle or hole of finite energy difference is present.
\end{abstract}

\vspace{2pc}
\maketitle

\section{Introduction}
The combination of Bethe ansatz (BA) and conformal field theory (CFT)
is a strong set of tools in the study of quantum mechanical systems in
one space dimension. To get insight into the correlations, a routinely
employed technique is to compute general expressions for correlation
asymptotes from CFT, fixing the critical exponents from the finite
size spectrum compared to the BA solution. This works well for static
correlations by taking the ground state as a reference state \cite{KorepinBOOK,1987_Bogoliubov_JPA_20}, while for
time-dependent correlations also contributions from certain impurity
configurations should be included
\cite{2009_Pereira_PRB_79,2012_Imambekov_RMP_84,2011_Kozlowski_JSTAT_P03018}.
This set of approaches thus provide a rather complete picture of (asymptotics of) equilibrium correlations in one-dimensional systems.

One of the outstanding benefits of the BA solution is the description
it provides of the full Hilbert space and the possibility to study
out-of-equilibrium states. The exact solvability can be attributed to the
existence of an infinite collection of local charges $\hat{Q}_n$ commuting with the
Hamiltonian $H$,
\begin{equation}
\label{eq:1}
 [H,\hat{Q}_n] = 0 ,\qquad  n\in\mathbb{N}.
\end{equation}
Out-of-equilibrium problems have attracted a lot of attention
recently regarding the question when and how unitary quantum systems
do or do not thermalize. Important in this respect is the idea that
correlations at late times can be computed in a generalized Gibbs
ensemble (GGE) \cite{2008_Rigol_NATURE_452,2007_Rigol_PRL_98} defined
not just by the Hamiltonian, but rather by all
\cite{2015_Ilievski_PRL_115,2015_Essler_PRA_91}  conserved (quasi) local
quantities
\begin{equation}
\label{eq:2}
  \hat{\rho}_{\rm GGE} = Z_{\rm GGE}^{-1}\exp\{-\sum_n \beta_n \hat{Q}_n\}.
\end{equation}
Equivalently, correlations can be computed on a single representative
eigenstate which can be determined by reasonings paralleling (generalizing) the
thermodynamic Bethe ansatz (gTBA) or by the Quench Action (QA)
method \cite{2013_Fagotti_PRB_87, 2012_Caux_PRL_109,
  2012_Essler_PRL_109, 2012_Mossel_JPA_45, 2011_Cassidy_PRL_106,
  2013_Caux_PRL_110,2013_Mussardo_PRL_111,
  2013_Pozsgay_JSTAT_P07003,2016_Alba_JSTAT_P043105,
  2013_Kormos_PRB_88, 2014_Fagotti_PRB_89, 2014_Wouters_PRL_113,
  2014_Pozsgay_PRL_113, 2014_Kormos_PRA_89,2014_DeNardis_PRA_89,
  2014_Sotiriadis_JSTAT_P07024,
  2015_Mestyan_JSTAT_P0400, 2015_Ilievski_PRL_115,2015_Essler_PRA_91, 2016_Caux_arxiv_1603.04689}. In the latter, one
constructs a free-energy functional straight from the overlaps of
the initial state with the eigenstates of the time evolution Hamiltonian
$H$ \cite{2013_Caux_PRL_110,2016_Caux_arxiv_1603.04689}. The GGE reasoning underscores the double role the Hamiltonian has
in equilibrium quantum mechanics in determining both the statistical
ensemble as well as the time evolution. Out of equilibrium, these two
roles are separated, at least in the presence of nontrivial local
conserved quantities.

A simple class of out-of-equilibrium states in BA solvable models
corresponds to the zero-temperature equivalent of a GGE with
nonmonotonic effective bare free energy (in gTBA language: driving
function). Such states can be specified by consecutive blocks of
filled quantum numbers in the Bethe ansatz solution, and in many ways
resemble the ground-state Fermi sea or a simple boosted version of it,
although now it combines several of such Fermi-sea blocks with
different mean momentum. It has been shown that even in such cases,
the description of correlation asymptotics is provided by multiple
CFTs and that the finite-size corrections to the spectrum can again be
used to obtain the critical exponents provided that the appropriate
GGE energy function $\epsilon_{GGE}(\lambda)$ is used
\cite{2013_Eriksson_JPA_46}.

The point of the present paper is to draw attention to a slightly
uncanny feature of the standard derivation of the finite-size spectrum
from Bethe ansatz
\cite{KorepinBOOK,2013_Eriksson_JPA_46,1987_Bogoliubov_JPA_20}, namely
that it requires the dressed energy function to vanish for excitations
at the Fermi points. This is done in equilibrium by substracting the
appropriate chemical potential. In other words, this requirement naturally follows when we use
the Hamiltonian that defines the statistical ensemble in a grand
canonical or GGE sense to measure energies, but out of equilibrium,
one may question the naturalness of this assumption. In particular,
when discussing dynamical correlations it is important to use the
time-evolution Hamiltonian to measure energies.  This suggests that
one should be able to derive the relation between critical exponents
and the finite-size spectrum for states of zero entropy density using
$H$---or any combination of the conserved quantities for that
matter---and the corresponding energy function, also when this is not
in line with the statistical ensemble.  This has indeed been verified
numerically in studies of dynamical correlations in out-of-equilibrium
zero entropy states in the Lieb-Liniger and XXZ models
\cite{2014_Fokkema_PRA_89, 2016_Vlijm_SCIPOST_inprep}.

We therefore revisit the derivation of the energy of zero-entropy
states and excitations in the limit of large system size and show that
many of the well known relations between the spectrum and CFT hold for
arbitray energy functions, but with essential modifications. In terms
of applications, the simplest example of such a situation occurs when
we choose to measure energies with respect to a different chemical
potential while still fixing a certain filling in a microcanonical
sense. This would of course change the energy of excitations, but
should not change the physics in an essential way. Another simple
application is that of a boosted state observed in the lab frame.  We
however here present a general treatment, applicable to any (multiply)
split Fermi sea in an integrable model.

% The simplest of a situation where the Hamiltonians for the ensemble
% and time evolution do not match is when we choose to
% measure energies with respect to a different chemical potential while still
% fixing a certain filling in a microcanonical sense. This would of course
% change the energy of excitations, but should not change the
% physics in an essential way. Another simple application is that of
% a boosted state observed in the lab frame. The main difference with the
% standard derivationis that the energy of particle excitations
% $\epsilon(\lambda)$ will not vanish at the generalized Fermi points of
% the distribution.

\section{Bethe ansatz and finite-size corrections}

To be specific, consider the repulsive Lieb-Liniger model defined by the
Hamiltonian
\begin{equation}
\label{eq:3}
H = \int dx \left[\partial_x\Psi^{\dag}(x)\partial_x\Psi(x) +c \Psi^{\dag}(x)\Psi^{\dag}(x)\Psi(x)\Psi(x) \right],\qquad c>0.
\end{equation}
The coordinate Bethe ansatz provides exact expressions for all eigenstates $\left|\{\lambda_j\}\right\rangle$ of the system with $N$ particles in a box
of size $L$ in terms of the rapidities $\lambda_j$ 
satisfying the Bethe equations \cite{KorepinBOOK}
\begin{equation}
\label{eq:4}
 L p_0(\lambda_j) + \sum_{k = 1}^N \theta(\lambda_j -\lambda_k) = 2\pi I_j.
\end{equation}
Here $p_0(\lambda) = \lambda$ is the bare momentum of particles and
$\theta(\lambda) = 2\arctan(\lambda/c)$ is the two-particle scattering
phase and $I_j$ are (half-odd) integers depending on whether $N$ is
(even) odd. All states are classified by specifying $N$ filled quantum
numbers $I_j$. The momentum and energy of a state are
\begin{equation}
\label{eq:5}
P = \sum_j \frac{2\pi}{L}I_j = \sum_j p_0(\lambda_j),\qquad
E = \sum_j \epsilon_0(\lambda_j)
\end{equation}
with $\epsilon_0(\lambda) = \lambda^2$. Note that the energy does not
include a chemical potential term and is really the eigenvalue of the
operator $H$. The conserved charges $\hat{Q}_n$ of the Lieb-Liniger model
can be taken to represent the monomials in the Bethe basis
\begin{equation}
\label{eq:6}
 \tilde{Q}\left|\{\lambda_j\}\right\rangle = Q_n \left|{\{\lambda_j\}}\right\rangle,\qquad Q_n = \sum_j \lambda_j^2
\end{equation}
such that $Q_0 = N$, $Q_1= P$ and $Q_2 = E$. Using the conserved
charges we can in principle construct a Hamiltonian for any
bare energy function $\epsilon_0(\lambda) = \sum_n \beta_n \lambda^n$
by matching the $\beta_n$ in the GGE.

Let us now consider a state $\left|{\{k_{ia}\}}\right\rangle$ which corresponds to
$n$ disjoint Fermi seas specified by left and right Fermi
momenta,
\begin{equation}
\label{eq:8}
k_{ia} ,\qquad a = R,L,\qquad i = 1,\ldots,n.
\end{equation}
These determine intervals of filled
quantum numbers between extrema $I_{ia} = (2\pi)^{-1}L k_{ia}$. 
We take the $I_{ia}$ to lie halfway between allowed quantum-number
slots such that the filled quantum numbers correspond to
\begin{equation}
\label{eq:9}
 \{I_j  \} = \bigcup_{i=1}^n  \{I_{iL}+ 1/2,I_{iL}+3/2,\ldots,I_{iR} - 1/2 \}.
\end{equation}
To take the thermodynamic limit $N,L\to \infty$ with $N/L$ fixed, we
introduce the rapidity density
\begin{equation}
\label{eq:10}
  \rho(\lambda_j) = \frac{1}{L(\lambda_{j+1} - \lambda_j)}.
\end{equation}
Using the Euler-Maclaurin formula, one can show that to order $1/L$
the density satisfies
\begin{equation}
\label{eq:rho}
 \rho(\lambda) = \frac{p_0'(\lambda)}{2\pi} + \sum_i \int_{\lambda_{iL}}^{\lambda_{iR}}\frac{d\nu}{2\pi}K(\lambda-\nu)\rho(\nu) + \frac{1}{24 L^2}\sum_{ia}\frac{s_aK'(\lambda - \lambda_{ia})}{2\pi \rho(\lambda_{ia})}
\end{equation}
where $K(\lambda) = \theta'(\lambda)$ and we introduced $s_{R,L} =
\pm 1$, and $\lambda_{ia}$ as the image of $I_{ia}$ in rapidity space
under the Bethe equations.
The energy similarly becomes (to order $1/L$)
\begin{equation}
\label{eq:E}
 E = L \sum_i \int_{\lambda_{iL}}^{\lambda_{iR}}d\lambda\, \epsilon_0(\lambda) \rho(\lambda) - \frac{1}{24 L}\sum_{ia} \frac{s_a \epsilon_0'(\lambda)}{\rho(\lambda_{ia})}.
\end{equation}
The remainder of this paper is largely concerned with the analysis of
these expressions.

We note that the solutions to other integrable models follow similar
lines with appropriate definitions of the functions
$\theta(\lambda),\, p_0(\lambda)$ and $\epsilon_0(\lambda)$.
For the XXZ model for instance, 
\begin{equation}
\label{eq:13}
 H = \sum_{j = 1}^L \left[S^x_jS^x_{j+1}+S^y_jS^y_{j+1} + \Delta (S^z_jS^z_{j+1} - 1/4) \right]
\end{equation}
with $\Delta = \cos \zeta \in (-1,1)$, we have
\begin{align}
\label{eq:14}
p_0(\lambda) &=2\arctan \left[\frac{\tanh(\lambda)}{\tan(\zeta/2)}\right],\qquad
\theta(\lambda) = 2 \arctan \left[\frac{\tanh(\lambda)}{\tan(\zeta)}\right]
\intertext{and}
\epsilon_0(\lambda) &= \frac{-2\sin^2 \zeta}{\cosh(2\lambda) -\cos \zeta}.
\end{align}
A complicating factor in XXZ is that solutions to the Bethe equations can be
complex. Using the string hypothesis the reasoning
can easily be generalized to these string states, but for simplicity we will
assume that quantum numbers and parameters are chosen such that we
deal with real rapidities. 
As is often the case in BA solvable models, the specific definitions do
not matter much in the later derivations, but the relations between
the functions do. This also means that 
$\epsilon_0(\lambda)$ can be chosen essentially at will.

\section{The energy of zero-entropy states}
\label{sec:energy}
Our first task is the evaluation of Eqs. (\ref{eq:rho}) and
(\ref{eq:E}). This follows standard
practice \cite{KorepinBOOK,2013_Eriksson_JPA_46}, but we include it for
completeness. We expand the solution to Eq. (\ref{eq:rho}) in powers of
$1/L$ as
\begin{equation}
\label{eq:15}
 \rho(\lambda) = \rho_{\infty}(\lambda) + \sum_{ia}\frac{s_a\rho_{ia}(\lambda)}{24 L^2 \rho_{\infty}(\lambda_{ia})}
\end{equation}
which results in the defining integral equations
\begin{align}
\label{eq:7}
  \rho_{\infty}(\lambda) & = \frac{p_0'(\lambda)}{2\pi} + \sum_{i} \int_{\lambda_{iL}}^{\lambda_{iR}}\frac{d\nu}{2\pi}K(\lambda-\nu)\rho_{\infty}(\nu),\\
  \rho_{ia}(\lambda) & = \frac{K'(\lambda-\lambda_{ia})}{2\pi} + \sum_{i} \int_{\lambda_{iL}}^{\lambda_{iR}}\frac{d\nu}{2\pi}K(\lambda-\nu)\rho_{ia}(\nu).
\end{align}
The equation for $\rho(\lambda)$ is the straightforward
generalization of the standard Lieb equation \cite{1963_Lieb_PR_130}.
The second equation shows that $\rho_{ia}(\lambda)$ is related to  the two-parameter function
$L(\lambda|\lambda')$ defined by
\begin{equation}
\label{eq:L}
 L(\lambda|\lambda') = \frac{K(\lambda-\lambda')}{2\pi} + \sum_i \int_{\lambda_{iL}}^{\lambda_{iR}}\frac{d\nu}{2\pi}K(\lambda-\nu)L(\nu|\lambda').
\end{equation}
Eq. (\ref{eq:L}) shows that, considered as integration kernels on the
domain specified by the Fermi rapidities $\lambda_{ia}$, the operator
$\widehat{(1 + L)}$ is the inverse of $\widehat{(1
  -\frac{K}{2\pi})}$.
Using this fact we obtain
\begin{equation}
\label{eq:ECFT}
 E = L \sum_i \int_{\lambda_{iL}}^{\lambda_{iR}}d\lambda \epsilon_0(\lambda)\rho_{\infty}(\lambda) - \sum_{ia}\frac{s_a \epsilon'(\lambda_{ia})}{24 L\rho_{\infty}(\lambda_{ia})}
\end{equation}
to order $1/L$, where the function $\epsilon(\lambda)$ is defined by
the integral equation
\begin{equation}
\label{eq:16}
 \epsilon(\lambda) = \epsilon_0(\lambda) +\sum_i\int_{\lambda_{iL}}^{\lambda_{iR}} \frac{d\nu}{2\pi}K(\lambda-\nu) \epsilon(\lambda).
\end{equation}
This definition is the direct analogue of the dressed energy function
in equilibrium settings which specifies the energy of the single
particle and hole excitations on the ground state, but, as we will see
later, this is not the case anymore. The true single particle
dispersion, which we will denote by $\tilde{\epsilon}(\lambda)$, will
in fact pick up additional contributions from the Fermi points
$\lambda_{ia}$ due to their nonzero energy.

CFT predicts a universal $1/L$ energy correction in terms of the
velocities of right and left moving modes of the form in 
Eq. (\ref{eq:ECFT}). However, here the velocity
\begin{equation}
\label{eq:17}
v_{ia} = \frac{\epsilon'(\lambda_{ia})}{2\pi \rho_{\infty}(\lambda_{ia})}
\end{equation}
differs from the dynamic velocity $\tilde{v}_{ia}$ from the dispersion
$\tilde{\epsilon}(\lambda)$ which governs the
propagation of correlations. 

From here on we will drop the subscript $\infty$ and denote by
$\rho(\lambda)$ the density in the thermodynamic limit.

\section{The shift function}
As it turns out, the shift function $F(\lambda|\lambda')$ determined
by the integral equation
\begin{equation}
\label{eq:F}
 F(\lambda|\lambda') = \frac{\theta(\lambda-\lambda')}{2\pi} + \sum_i \int_{\lambda_{iL}}^{\lambda_{iR}}\frac{d\nu}{2\pi}K(\lambda-\nu)F(\nu|\lambda')
\end{equation}
plays an important role. Its definition can be obtained by considering
a particle-hole excitation with rapidity $\lambda_p$ for the particle
and $\lambda_h$ for the hole, as is discussed in standard textbooks
\cite{KorepinBOOK}. Denoting $\lambda_j$ for the solution of
the Bethe equations for the state $\left|{\{k_{ia}\}}\right\rangle$ and
$\tilde{\lambda}_j$ for the excited state, we can define the shift
function for the particle-hole excitation as
\begin{equation}
\label{eq:19}
 F(\lambda_j|\lambda_p,\lambda_h) = \frac{\lambda_j - \tilde{\lambda}_j}{\lambda_{j+1} - \lambda_j}.
\end{equation}
From the Bethe equations one can show that
$F(\lambda|\lambda_p,\lambda_h) = F(\lambda|\lambda_p) -
F(\lambda|\lambda_h)$ with definitions according to (\ref{eq:F}).  

In this section we collect various results on the shift function for
zero entropy states which
are quite  useful. Especially for the case of the ground state this is
all known, but a discussion of the generality seems unavailable in the
literature or is at
least hard to find.

Since $\partial_{\lambda'}\theta(\lambda-\lambda') =
-K(\lambda-\lambda')$ we easily see that
\begin{equation}
\label{eq:20}
 \partial_{\lambda'}F(\lambda|\lambda') = - L(\lambda|\lambda').
\end{equation}
It is worth noting that $L(\lambda|\lambda') = L(\lambda'|\lambda)$,
but 
\begin{equation}
\label{eq:dlamF}
 \partial_{\lambda} F(\lambda|\lambda') = L(\lambda|\lambda') - \sum_{ia}s_aL(\lambda|\lambda_{ia})F(\lambda_{ia}|\lambda')
\end{equation}
which follows from Eq. (\ref{eq:F}) by using a partial integration.
Another, very useful, relation is
\begin{equation}
\label{eq:Frel}
 F(\lambda|\lambda') + F(\lambda'|\lambda) = \sum_{ia}s_aF(\lambda_{ia}|\lambda)F(\lambda_{ia}|\lambda').
\end{equation}
%%%%%%%%%%%%%%%%%%
% This is derived as follows:
% \begin{align}
% \label{eq:63}
%   F(\lambda'|\lambda) =& \frac{\theta(\lambda'-\lambda)}{2\pi} + \sum_i\int_{\lambda_{iL}}^{\lambda_{iR}}\frac{d\nu}{2\pi}\theta'(\lambda'-\nu)F(\nu|\lambda)\\
% =& - \frac{\theta(\lambda-\lambda')}{2\pi} - \sum_i \int_{\lambda_{iL}}^{\lambda_{iR}} \frac{d\nu}{2\pi}L(\lambda|\nu)\theta(\nu-\lambda')\\
% &+\sum_{ia}s_a\left[\frac{\theta(\lambda_{ia} - \lambda')}{2\pi} + \sum_j\int_{\lambda_{jL}}^{\lambda_{jR}}\frac{d\nu}{2\pi}L(\lambda_{ia}|\nu)\theta(\nu-\lambda')   \right]F(\lambda_{ia}|\lambda)\\
%   =& -F(\lambda|\lambda') + \sum_{ia}s_aF(\lambda_{ia}|\lambda)F(\lambda_{ia}|\lambda')
% \end{align}
%%%%%%%%%%%%%%%%%
This in particular implies
\begin{equation}
\label{eq:Finv}
 \sum_{kc}[\delta_{ia,kc} - s_cF(\lambda_{kc}|\lambda_{ia})][\delta_{jb,kc} - s_bF(\lambda_{kc}|\lambda_{jb})] = \delta_{ia,jb}
\end{equation}
hence we have found a matrix-inverse pair
\begin{equation}
\label{eq:M}
U_{ia,jb} = \delta_{ia,jb} - s_bF(\lambda_{jb}|\lambda_{ia}),\qquad
[U^{-1}]_{ia,jb} = \delta_{ia,jb} - s_b F(\lambda_{ia}|\lambda_{jb}).
\end{equation}
Finally,
\begin{equation}
\label{eq:25}
  \sum_{ia} s_a F(\lambda|\lambda_{ia})F(\lambda'|\lambda_{ia})   = \sum_{ia}s_a F(\lambda_{ia}|\lambda)F(\lambda_{ia}|\lambda').
\end{equation}
Strictly speaking, the function $F(\lambda|\lambda')$ does not encode
the shift of rapidities when a single particle or hole is created in
bosonic models such as Lieb-Liniger and XXZ due to the $1/2$ shift in
the quantum number lattice when we change particle-number parity. Rather,
$F(\lambda|\lambda')$ represents the phase shifts in the fermionic
dual which is the Cheon-Shigehara model
\cite{1998_Cheon_PLA_243,1999_Cheon_PRL_82} for Lieb-Liniger and spinless
lattice fermions for XXZ.

Using the conventions that adding a particle shifts the occupied
quantum numbers to the left while adding a hole shifts them to the
right, we can define the bosonic shift function
\begin{equation}
\label{eq:24}
 F_B(\lambda|\lambda') = \frac{\theta(\lambda-\lambda')-\pi}{2\pi} + \sum_i \int_{\lambda_{iL}}^{\lambda_{iR}}\frac{d\nu}{2\pi}K(\lambda-\nu)F_B(\nu|\lambda').
\end{equation}
The relation between $F(\lambda|\lambda')$ and $F_B(\lambda|\lambda')$
may be expressed as
\begin{equation}
\label{eq:28}
 F_B(\lambda|\lambda') = F(\lambda|\lambda') -\frac{1}{2}Z(\lambda)
\end{equation}
where $Z(\lambda)$ is the analogue of the dressed charge
\begin{equation}
\label{eq:31}
 Z(\lambda) = 1 +\sum_i \int_{\lambda_{iL}}^{\lambda_{iR}}\frac{d\nu}{2\pi}K(\lambda-\nu)Z(\nu)
\end{equation}
which is related to critical exponents in the case of the ground
state, but a similar interpretation is not present in the general
case. Note that derivatives with respect to $\lambda$ or $\lambda'$
of $F(\lambda|\lambda')$ and $F_B(\lambda|\lambda')$ coincide.

From hereon we will implicitly assume that we deal with the fermionic
version of the models. This makes the connection with the (fermionic)
effective field theory most transparent. The difference is only
important for single particle or hole excitations.

\section{Energy and momentum of excitations}
In order to determine the single-particle dispersion function, let us
consider again a particle-hole excitation on top of the state
$\left|\{k_{ia}\}\right\rangle$ with particle rapidity $\lambda_p$ and hole
rapidity $\lambda_h$. Let $\lambda_j$ and $\tilde{\lambda}_j$ again denote
the solution to the Bethe equations before and after excitation. The
energy difference
\begin{equation}
\label{eq:26}
 \Delta E(\lambda_p,\lambda_h) = \epsilon_0(\lambda_p) - \epsilon_0(\lambda_h) +\sum_j[\epsilon_0(\tilde{\lambda}_j) - \epsilon_0(\lambda_j)]
\end{equation}
can be expressed in the thermodynamic limit as
\begin{equation}
\label{eq:27}
 \Delta E(\lambda_p,\lambda_h) = \tilde{\epsilon}(\lambda_p) - \tilde{\epsilon}(\lambda_h)
\end{equation}
with
\begin{equation}
\label{eq:epstil0}
  \tilde{\epsilon}(\lambda) = \epsilon_0(\lambda) - \sum_i\int_{\lambda_{iL}}^{\lambda_{iR}}d\nu\, \epsilon_0'(\nu)F(\nu|\lambda).
\end{equation}
By a partial integration we obtain
\begin{equation}
\label{eq:29}
 \tilde{\epsilon}(\lambda) = \epsilon_0(\lambda) - \sum_{ia}s_a\epsilon_0(\lambda_{ia})F(\lambda_{ia}|\lambda) +\sum_i \int_{\lambda_{iL}}^{\lambda_{iR}}d\nu\, \epsilon_0(\nu) \partial_{\nu}F(\nu|\lambda)
\end{equation}
from where Eq. (\ref{eq:dlamF}) expresses the actual single-particle
dispersion as
\begin{equation}
\label{eq:epstil}
 \tilde{\epsilon}(\lambda) = \epsilon(\lambda) - \sum_{ia}s_a \epsilon(\lambda_{ia})F(\lambda_{ia}|\lambda).
\end{equation}
Note that this indeed differs from $\epsilon(\lambda)$ when
$\epsilon(\lambda_{ia})\neq 0$ and we have nontrivial backflow $F(\lambda_{ia}|\lambda)\neq 0$.

The momentum of a particle is defined by the equation
\begin{equation}
\label{eq:k}
 k(\lambda) = p_0(\lambda) -\sum_i \int_{\lambda_{iL}}^{\lambda_{iR}}d\nu\, p_0'(\nu)F(\nu|\lambda)
\end{equation}
from which it is easy to see that $k'(\lambda) = 2\pi
\rho(\lambda)$ as in the equilibrium case. 

Of particular interest is the energy and velocity of particles close
to the Fermi points $k_{ia}$. We note that Eqs. (\ref{eq:epstil}) and
(\ref{eq:Finv}) imply
\begin{align}
\label{eq:32}
  \tilde{\epsilon}(\lambda_{ia}) &= \sum_{jb}[\delta_{ia,jb} - s_bF(\lambda_{jb}|\lambda_{ia})]\epsilon(\lambda_{jb}),\\
  \label{eq:epstilia}
  \epsilon(\lambda_{ia}) &= \sum_{jb}[\delta_{ia,jb} - s_bF(\lambda_{ia}|\lambda_{jb})]\tilde{\epsilon}(\lambda_{jb}).
\end{align}
The Fermi velocity for the Fermi point $k_{ia}$ is defined as
\begin{equation}
\label{eq:33}
 \tilde{v}_{ia} = \left.\frac{\partial \varepsilon}{\partial k}\right|_{k=k_{ia}} = \frac{\tilde{\epsilon}'(\lambda_{ia})}{2\pi \rho(\lambda_{ia})}.
\end{equation}
The relation between $\epsilon(\lambda)$ and
$\tilde{\epsilon}(\lambda)$ can also be expressed as
\begin{equation}
\label{eq:34}
 \epsilon(\lambda) = \tilde{\epsilon}(\lambda) - \sum_{ia}s_a \tilde{\epsilon}(\lambda_{ia})F(\lambda|\lambda_{ia}).
\end{equation}

\section{Finite-size spectrum and critical exponents}
Now that we have established the energy of zero-entropy states
to order $1/L$ [Eq. (\ref{eq:E})] and the energy of particle and hole
excitations in the thermodynamic limit [Eq. (\ref{eq:epstil})] we will
ask the usual question: what is the change in energy upon adding or
removing particles very close to the Fermi points $I_{ia}$? Let us
consider a state defined by $I_{ia} \to I_{ia} + s_aN_{ia}$,
i.e. $N_{ia}$ denotes the number of particles added or removed at the
Fermi point $k_{ia}$.

In terms of the quantum numbers 
\begin{equation}
\label{eq:ND}
N_i = L \int_{\lambda_{iL}}^{\lambda_{iR}}d\lambda\, \rho(\lambda),\qquad
D_i =L\left\{\int_{-\infty}^{\lambda_{iL}} - \int_{\lambda_{iR}}^{\infty}\right\}d\lambda\,\rho(\lambda)
\end{equation}
we have
\begin{equation}
\label{eq:18}
  N_{ia} = \frac{\Delta N_i + s_a \Delta D_i}{2}
\end{equation}
where $\Delta N_i,\, \Delta D_i$ denotes the change in $N_i,\, D_i$ .
We can also express the variation of the state in terms of the change in the Fermi
rapidities $\lambda_{jb} \to \lambda_{jb} +\delta \lambda_{jb}$. The
definitions in Eq. (\ref{eq:ND}) allow us to compute the Jacobian
\begin{equation}
\label{eq:23}
 \frac{\partial N_{ia}}{\partial \lambda_{jb}} = s_aL\left\{\rho(\lambda_{ia})\delta_{ia,jb} + \frac{1}{2} \int_{-\infty}^{\infty}d\lambda\, s_a \mathrm{sgn}(\lambda_{ia}-\lambda)\frac{d \rho}{d \lambda_{jb}}(\lambda) \right\}.
\end{equation}
Using that
\begin{equation}
\label{eq:30}
\frac{\partial \rho}{\partial \lambda_{jb}}(\lambda) = s_b\rho(\lambda_{jb})L(\lambda_{jb}|\lambda)
=- s_b\rho(\lambda_{jb})\partial_{\lambda}F(\lambda_{jb}|\lambda)
\end{equation}
and a partial integration one finds
\begin{equation}
\label{eq:36}
 \frac{\partial N_{ia}}{\partial \lambda_{jb}} = Ls_b \rho(\lambda_{jb})[\delta_{ia,jb} - s_aF(\lambda_{jb}|\lambda_{ia})].
\end{equation}
We recognize the matrix $[U^{-1}]_{jb,ia}$ from Eq. (\ref{eq:M}), which
immediately gives
\begin{equation}
\label{eq:37}
 \frac{\partial \lambda_{ia}}{\partial N_{jb}} = \frac{1}{L s_a \rho(\lambda_{ia})}[\delta_{ia,jb} - s_aF(\lambda_{ia}|\lambda_{jb})]
\end{equation}
and therefore we can express
\begin{equation}
\label{eq:38}
 \delta \lambda_{ia} = \sum_{jb} \frac{\delta_{ia,jb}-s_aF(\lambda_{ia}|\lambda_{jb})}{L s_a \rho(\lambda_{ia})}N_{jb}.
\end{equation}
Since the Fermi momenta are directly related to the numbers $I_{ia}$,
the change in Fermi momentum is
\begin{equation}
\label{eq:delk}
 \delta k_{ia} = \frac{L s_aN_{ia}}{2\pi}
\end{equation}
which can also be obtained from the definition of $k(\lambda)$ in Eq. (\ref{eq:k}).
Hence also the relations
\begin{align}
\label{eq:39}
  \frac{\partial k_{ia}}{\partial \lambda_{jb}} &= [\delta_{ia,jb} - s_bF(\lambda_{jb}|\lambda_{ia})]2\pi \rho(\lambda_{jb}),\\
  \label{eq:dlamdk}
\frac{\partial \lambda_{ia}}{\partial k_{jb}} &= \frac{1}{2\pi \rho(\lambda_{ia})}[\delta_{ia,jb} - s_b F(\lambda_{ia}|\lambda_{jb})]
\end{align}
are valid.

Let us consider corrections to the energy $E$ in Eq. (\ref{eq:E}) to
order $1/L$ when $k_{ia} \to k_{ia} + \delta k_{ia}$. We express
\begin{align}
\label{eq:40}
  \delta E &= \sum_{ia} \frac{\partial E}{\partial \lambda_{ia}}\delta\lambda_{ia} + \frac{1}{2}\sum_{ia,jb}\frac{\partial^2 E}{\partial \lambda_{ia}\partial \lambda_{jb}}\delta\lambda_{ia}\delta\lambda_{jb}\\
  \intertext{or equivalently}
\delta E&= \sum_{ia} \frac{\partial E}{\partial k_{ia}}\delta k_{ia} + \frac{1}{2}\sum_{ia,jb}\frac{\partial^2 E}{\partial k_{ia}\partial k_{jb}}\delta k_{ia}\delta k_{jb}.
\end{align}
Note that these corrections can only come from the extensive
contribution to $E$ since $\delta \lambda_{ia}$ and $\delta k_{ia}$ are
of order $1/L$. 

From Eq. (\ref{eq:E}) and (\ref{eq:epstilia}) we obtain
\begin{equation}
\label{eq:41}
 \frac{\partial E}{\partial \lambda_{ia}} = Ls_a\rho(\lambda_{ia}) \epsilon(\lambda_{ia}) = \sum_{jb}L s_a\rho(\lambda_{ia})[\delta_{ia,jb} - s_b F(\lambda_{ia}|\lambda_{jb})]\tilde{\epsilon}(\lambda_{jb})
\end{equation}
which together with Eqs. (\ref{eq:dlamdk}) shows
\begin{equation}
\label{eq:42}
  \frac{\partial E}{\partial k_{ia}} = \frac{s_aL}{2\pi}\tilde{\epsilon}(\lambda_{ia})
\end{equation}
so that
\begin{equation}
\label{eq:43}
 \delta E^{(1)} = \sum_{ia}\frac{\partial E}{\partial k_{ia}}\delta k_{ia} = \sum_{ia}\tilde{\epsilon}(\lambda_{ia})N_{ia}
\end{equation}
(where we have introduced the notation $\delta E^{(n)}$ for the order
$L^{-n}$ term in $\delta E$).

Next, consider the second order correction
\begin{equation}
\label{eq:47}
 \delta E^{(2)} = \frac{1}{2}\sum_{ia,jb}\frac{\partial E}{\partial k_{ia} \partial k_{jb}} \delta k_{ia}\delta k_{jb}.
\end{equation}
From Eq. (\ref{eq:epstil0}) we find that
\begin{equation}
\label{eq:44}
 \frac{\partial \tilde{\epsilon}}{\partial \lambda_{jb}}(\lambda) = -s_b \epsilon_0'(\lambda_{jb})F(\lambda_{jb}|\lambda) -\sum_i\int_{\lambda_{iL}}^{\lambda_{iR}}d\nu\, \epsilon_0'(\nu) \frac{\partial F}{\partial \lambda_{jb}}(\nu|\lambda),
\end{equation}
which together with
\begin{equation}
\label{eq:45}
 \frac{\partial F}{\partial \lambda_{jb}}(\lambda|\lambda') = s_b L(\lambda|\lambda_{jb})F(\lambda_{jb}|\lambda')
\end{equation}
can be used to show that
\begin{equation}
\label{eq:46}
  \frac{\partial \tilde{\epsilon}}{\partial \lambda_{jb}}(\lambda) = -s_b \tilde{\epsilon}'(\lambda_{jb})F(\lambda_{jb}|\lambda).
\end{equation}
For the derivation it is useful to note
\begin{align}
\label{eq:11}
  \epsilon'(\lambda) &= \epsilon_0'(\lambda)+\sum_i\int_{\lambda_{iL}}^{\lambda_{iR}}d\nu\, \partial_{\lambda}L(\lambda|\nu)\epsilon_0(\nu),\\
  \tilde{\epsilon}'(\lambda) & = \epsilon'(\lambda) + \sum_{ia}s_a\epsilon(\lambda_{ia})L(\lambda_{ia}|\lambda).
\end{align}
%% %%%%%%%%%%%
% This derivation follows from
% \begin{multline}
% \label{eq:12}
%   \frac{\partial \tilde{\epsilon}}{\partial \lambda_{jb}}(\lambda) = -s_b\left[ \epsilon_0'(\lambda_{jb})+ \sum_{ia}s_a\epsilon_0(\lambda_{ia})L(\lambda_{ia}|\lambda_{jb})\right]F(\lambda_{jb}|\lambda)\\
%                                                                      +s_b\sum_i \int_{\lambda_{iL}}^{\lambda_{iR}}d\nu\, \epsilon_0(\nu) \partial_{\nu}L(\nu|\lambda_{jb})F(\lambda_{jb}|\lambda)
% \end{multline}
% using that
% \begin{equation}
% \label{eq:21}
%  \partial_{\nu}L(\nu|\nu') = -\partial_{\nu}\partial_{\nu'}F(\nu|\nu') = \partial_{\nu'}L(\nu'|\nu) - \sum_{ia}s_a L(\nu|\lambda_{ia})L(\lambda_{ia}|\nu').
% \end{equation}
% %%%%%%%%%%%%%%%%%
Computing
\begin{equation}
\label{eq:22}
 \frac{\partial}{\partial \lambda_{jb}}\left(\frac{\partial E}{\partial k_{ia}}\right) = \frac{s_aL}{2\pi}[\delta_{ia,jb} - s_bF(\lambda_{jb}|\lambda_{ia})]\tilde{\epsilon}'(\lambda_{jb})
\end{equation}
we thus find
\begin{align}
\label{eq:48}
\delta E^{(2)} &= \frac{1}{2}\sum_{ia,jb,kc} \frac{\partial \lambda_{kc}}{\partial k_{jb}} \frac{\partial }{\partial \lambda_{kc}}\left(\frac{\partial E}{\partial k_{ia}}\right)\\
&= \frac{1}{L}\sum_{ia,jb,kc}\frac{\tilde{\epsilon}'(\lambda_{kc})}{2 \rho(\lambda_{kc})}[\delta_{ia,kc}- s_c F(\lambda_{kc}|\lambda_{ia})][\delta_{jb,kc}- s_c F(\lambda_{kc}|\lambda_{jb})]N_{ia}N_{jb}.
\end{align}
Now it is easy to also incorporate the number of particle-hole
excitations corresponding to a total number of mementum quanta $n_{ia}$ close to
the Fermi point $k_{ia}$ and arrive at the general result for the spectrum
\begin{equation}
\label{eq:delE}
 \delta E = \sum_{ia} \tilde{\epsilon}(\lambda_{ia})N_{ia} + \frac{2\pi}{L}\sum_{ia}s_a\tilde{v}_{ia}\left[n_{ia} + \frac{1}{2}\left(\sum_{jb}U_{jb,ia}N_{jb}. \right)^2\right]
\end{equation}
This is valid for general zero-entropy states $\left|\{k_{ia}\}\right\rangle$ and general energy
functions  $\epsilon_0(\lambda)$ with
\begin{equation}
\label{eq:50}
U_{ia,jb} = \delta_{ia,jb} - s_bF(\lambda_{ia}|\lambda_{jb}), \qquad [U^{-1}]_{ia,jb} = s_as_bU_{jb,ia}.
\end{equation}
Note that the velocity $s_a\tilde{v}_{ia}$ can be negative in the
current setup.

The matrix $U_{ia,jb}$ is identified with the matrix of the Bogoliubov
transformation diagonalizing the
multi-component Tomonaga-Luttinger Hamiltonian describing the state 
\cite{2016_Vlijm_SCIPOST_inprep}. These parameters
determine the exponents of critical correlations, i.e. the conformal
dimensions of scaling fields in the language of CFT.

\section{The symmetric case}
In the case of a symmetric quantum number configuration,
$I_{iL} = - I_{n+1-iR}$, we have the equalities
\begin{equation}
\label{eq:52}
\tilde{v}_{iL} = - \tilde{v}_{n+1-iR}\qquad
\text{and}\qquad U_{ia,jb} = U_{n+1-i\bar{a},n+1-j\bar{b}}
\end{equation}
(with $\bar{L}=R$ and
$\bar{R}=L$). Define the matrices
\begin{align}
\label{eq:55}
  Z_{ij} &= U_{iR,jR} - U_{n+1-iL,jR} &&= \delta_{ij} - F(\lambda_{jR}|\lambda_{iR}) + F(\lambda_{jR}|\lambda_{n+1-iL}),\\
  Y_{ij} &= U_{iR,jR} + U_{n+1-iL,jR} &&= \delta_{ij}- F(\lambda_{jR}|\lambda_{iR}) - F(\lambda_{jR}|\lambda_{n+1-iL}).
\end{align}
Using that in the symmetric case $F(-\lambda|-\lambda') =
-F(\lambda|\lambda')$ and $\lambda_{iL} = - \lambda_{n+1-iR}$,
Eq. (\ref{eq:Frel}) gives
\begin{equation}
\label{eq:56}
 \sum_k Z_{ik}Y_{jk} = \delta_{ij}
\end{equation}
and so $Z^{-1} = Y^T$ which is closely related to the general relation
$[U^{-1}]_{ia,jb} = s_as_b U_{jb,ia}$.

The finite-size correction to the energy can then be written as
\begin{equation}
\label{eq:58}
\delta E = \sum_{i} \tilde{\epsilon}_i \tilde{N}_i + \frac{2\pi}{L}\sum_i \frac{\tilde{v}_i}{2}\left[\left(\sum_j [Z^{-1}]_{ij} \tilde{N}_j\right)^2 +\left( \sum_j Z_{ji}  \tilde{D}_j\right)^2\right] 
\end{equation}
where $\tilde{\epsilon}_i=\tilde{\epsilon}(\lambda_{iR})$,
$\tilde{v}_i = \tilde{v}_{iR}$ and
\begin{equation}
\label{eq:59}
\tilde{N}_i = N_{iR} + N_{n+1-iL},\qquad \tilde{D}_i = N_{iR} -  N_{n+1-iL}.
\end{equation}
We can write
\begin{equation}
\label{eq:Z}
 Z_{ij} = \delta_{ij} + \int_{\lambda_{n+1-iL}}^{\lambda_{iR}}d\nu\, F(\lambda_{jR}|\nu).
\end{equation}
We can also obtain this matrix from as $Z_{ij} =
\xi_{ij}(\lambda_{jR})$ where $\xi_{ij}(\lambda)$ is defined by
\begin{equation}
\label{eq:61}
 \xi_{ij}(\lambda) = \delta_{ij} + \sum_k \int_{\lambda_{n+1-iL}}^{\lambda_{iR}}\frac{d\nu}{2\pi}K(\lambda-\nu)\xi_{kj}(\nu)
\end{equation}
which is straightforward to derive using the relation
$\partial_{\lambda'}F(\lambda|\lambda') = -L(\lambda|\lambda')$ from
Eq. (\ref{eq:Z}). Hence, in the symmetric case we reach the same
conclusion as Ref. \cite{2013_Eriksson_JPA_46}, namely that the
critical exponents can equivalently be expressed in terms of a dressed
charge matrix $\xi_{ij}(\lambda)$ similar to models solvable by nested
Bethe ansatz
\cite{1989_Izergin_JPA_22,1990_Tsvelik_PRB_42,1992_Frahm_JPA_25,1987_Pokrovsky_ZETF_93,
  1997_Frahm_JPA_30,1990_Frahm_JPA_23,2000_Zvyagin_LTP_26,2001_Zvyagin_EPJB_19,2003_Zvyagin_PRB_68,1990_Frahm_PRB_42,HubbardBOOK}.

\section{Impurity configurations}

Let us consider an impurity configuration defined by one hole with
$\lambda_h$ in, or one particle with $\lambda_p$ outside of one of the
Fermi-sea blocks and ask again what the spectrum of excitations at the
Fermi points is to order $1/L$.  Here, the energy of the state
$\left|{\{k_{ia}\}}\right\rangle$ still serves as the reference. We restrict the
analysis to the particle case, as the case of a hole just introduces
appropriate minus signs. Note that we assume to work in the fermionic
dual here such that $F(\lambda|\lambda')$ encodes the shift of
rapidities for a single-particle excitation.

In the case of an impurity we have to go back to the derivation of for
the root density in Sec. \ref{sec:energy} to order $1/L^2$. From the
Bethe equations we find
\begin{equation}
\label{eq:35}
 \rho(\lambda) = \frac{p_0'(\lambda)}{2\pi} + \sum_i\int_{\lambda_{iL}}^{\lambda_{iR}}\frac{d\nu}{2\pi}K(\lambda-\nu)\rho(\nu) + \frac{K(\lambda-\lambda_p)}{2\pi L} +\frac{1}{24 L^2} \sum_{ia}\frac{s_aK'(\lambda-\lambda_{ia})}{2\pi\rho(\lambda_{ia})}
\end{equation}
in this case. The solution for $\rho(\lambda)$ thus has an extra
contribution due to the impurity
\begin{equation}
\label{eq:51}
 \rho(\lambda) = \rho_{\infty}(\lambda) + \frac{\rho_{\mathrm{imp}}(\lambda|\lambda_p)}{L} + \sum_{ia} \frac{\rho_{ia}(\lambda)}{24 L^2\rho_{\infty}(\lambda_{ia})}
\end{equation}
where clearly
\begin{equation}
\label{eq:53}
 \rho_{\mathrm{imp}}(\lambda|\lambda_p) = L(\lambda|\lambda_p).
\end{equation}
Going back to the definitions of $N_i$ and $D_i$, we find that
\begin{equation}
\label{eq:Niimp}
N_i = n^{\rm imp}_i + L \int_{\lambda_{iL}}^{\lambda_{iR}}d\lambda\, \rho_{\infty}(\lambda),\qquad
D_i = d^{\rm imp}_i + L\left\{ \int_{-\infty}^{\lambda_{iL}} -\int_{\lambda_{iR}}^{\infty}\right\}  d\lambda\, \rho_{\infty}(\lambda)
\end{equation}
with
\begin{align}
\label{eq:64}
n_i^{\rm imp} &= \int_{\lambda_{iL}}^{\lambda_{iR}}d\lambda\, L(\lambda|\lambda_p) = -F(\lambda_{iR}|\lambda_p) + F(\lambda_{iL}|\lambda_p),\\
 d_i^{\rm imp} &= \left\{ \int_{-\infty}^{\lambda_{iL}} -\int_{\lambda_{iR}}^{\infty}\right\}d\lambda\, L(\lambda|\lambda_p) = - F(\lambda_{iR}|\lambda_p) - F(\lambda_{iR}|\lambda_p).
\end{align}
Considering the energy difference of the state $\left|{\{k_{ia}\}}\right\rangle$  and
the state defined by the
addition of particles at the Fermi points according to the numbers
$\{N_{ia}\}$  and the additional particle impurity with quantum number
$I_p$ leads to
\begin{equation}
\label{eq:65}
\delta E  = \tilde{\epsilon}(\lambda_p) + \sum_{ia}\tilde{\epsilon}(\lambda_{ia})[N_{ia}-n^{\rm imp}_{ia}] 
+ \frac{2\pi}{L}\sum_{ia}s_a\tilde{v}_{ia}\left[n_{ia} + \frac{1}{2}\left(\sum_{jb}U_{jb,ia}[N_{jb} -n^{\rm imp}_{jb}] \right)^2\right]
\end{equation}
with
\begin{equation}
\label{eq:66}
 n^{\rm imp}_{ia} = \frac{n_i^{\rm imp}+ s_ad_i^{\rm imp}}{2} = - s_aF(\lambda_{ia}|\lambda_p)
\end{equation}
which follows by the same reasoning as leading up to
Eq. (\ref{eq:delE}) but using  Eq. (\ref{eq:Niimp}).

% Defining $\mathcal{N}_{ia}$ to be given by Eqs. such that
% \begin{equation}
% \label{eq:60}
%   \mathcal{N}_{ia} = L \int_{-\infty}^{\infty}d\lambda\, s_a \frac{1}{2}\sgn(\lambda-\lambda_{ia})\rho(\lambda) = N_{ia} +  \int_{-\infty}^{\infty}d\lambda\, s_a \frac{1}{2}\sgn(\lambda-\lambda_{ia})L(\lambda|\lambda_{ia})
% \end{equation}
% in the presence of the impurity and 
% without changing the values of $\lambda_{ia}$ gives
% \begin{equation}
% \label{eq:54}
%  \mathcal{N}_{ia} = N_{ia} + s_a F(\lambda_{ia}|\lambda_p).
% \end{equation}
% The derivation for the $1/L$ spectrum now goes through by replacing
% $N_{ia}$ by $\mathcal{N}_{ia}$, so we find
% \begin{multline}
% \label{eq:57}
%  \delta E = \sum_{ia}\tilde{\epsilon}(\lambda_{ia})[N_{ia} -
%  s_aF(\lambda_{ia}|\lambda_p)] \\
%  +\frac{2\pi}{L}\sum_{ia}s_a\tilde{v}_{ia}\left[n_{ia} + \frac{1}{2}\left(\sum_{jb}[\delta_{ia,jb}-s_aF(\lambda_{ia}|\lambda_{jb})][N_{jb} -s_b F(\lambda_{jb}|\lambda_p)] \right)^2 \right]
% \end{multline}
A hole impurity just replaces $F(\lambda_{ia}|\lambda_p) \to -
F(\lambda_{ia}|\lambda_h)$. The generalization to multiple impurities
is straightforward. 

\section{Conclusion}

We have considered the energy of excitations on states of zero entropy
density in the Lieb-Liniger and other Bethe ansatz solvable
models. These states can be considered as the zero-temperature limit
of a statistical ensemble defined by a generalized Hamiltonian in the
spirit of the GGE. We explicitly allowed the energies to be measured
with a different Hamiltonian which generically would correspond to the
physical Hamiltonian of the model. We have shown that the dispersion
function is not necessarily determined by a single integral equation,
but includes contributions from the generalized Fermi points that may
have finite energy in the situation under consideration. We derived a
generalization of the expression for finite-size corrections to the
spectrum.  This derivation is valid for arbitrary bare energy
functions $\epsilon_0(\lambda)$ constructed from the eigenvalues of
local charges on the Bethe basis and also for arbitrary configurations
of Fermi seas. The energy corrections related to addition or
subtraction of particles at the generalized Fermi points, which are
directly related to critical exponents, are expressed in terms of the
shift function and only for a symmetric configuration can this be
expressed in terms of a dressed charge matrix. Similar expressions are
derived in the presence of an additional particle and hole impurity.

Our results are interesting in the light of recent developments in the
correspondence between Bethe ansatz solvable models and effective
field theory methods. The characteristic power-law behavior of
correlations well known from the correspondence with CFT can be
interpreted in terms of the Anderson orthogonality catastrophe due to
the phase shift of the modes at the Fermi points. While for static
correlations one only considers Umklapp-like configurations, time
dependent correlations include additional contributions from certain
impurity configurations, but the logic in both cases is remarkably
similar. The point is that the power law exponents are completely
determined by the phase shifts (static data) while the characteristic
frequencies of oscillations in space and time are determined by the
momentum and energy differences of the reference state with the
Umklapp and impurity correlations. Our work suggests that this
decomposition of effects can be extended to out-of-equilibrium
correlations of zero-entropy states and the power-law exponents depend
only on the scattering data of the theory and are Hamiltonian
independent.

\section*{Acknowledgements}
We gladfully thank Yuri van Nieuwkerk for helpful discussions.
This work is part of the research programme of the Foundation for
Fundamental Research on Matter (FOM), which is part of the Netherlands
Organisation for Scientific Research (NWO).

\newpage
\section*{References}
\bibliographystyle{iopart-num}
\bibliography{ise_bibtex_library,extra}

\providecommand{\newblock}{}
\begin{thebibliography}{10}
\expandafter\ifx\csname url\endcsname\relax
  \def\url#1{{\tt #1}}\fi
\expandafter\ifx\csname urlprefix\endcsname\relax\def\urlprefix{URL }\fi
\providecommand{\eprint}[2][]{\url{#2}}

\bibitem{KorepinBOOK}
Korepin V~E, Bogoliubov N~M and Izergin A~G 1993 {\em Quantum Inverse
  Scattering Method and Correlation Functions\/} (Cambridge Univ. Press)

\bibitem{1987_Bogoliubov_JPA_20}
Bogoliubov N~M, Izergin A~G and Reshetikhin N~Y 1987 {\em Journal of Physics A:
  Mathematical and General\/} {\bf 20} 5361
  \urlprefix\url{http://stacks.iop.org/0305-4470/20/i=15/a=047}

\bibitem{2009_Pereira_PRB_79}
Pereira R~G, White S~R and Affleck I 2009 {\em Phys. Rev. B\/} {\bf 79}(16)
  165113 \urlprefix\url{http://link.aps.org/doi/10.1103/PhysRevB.79.165113}

\bibitem{2012_Imambekov_RMP_84}
Imambekov A, Schmidt T~L and Glazman L~I 2012 {\em Rev. Mod. Phys.\/} {\bf
  84}(3) 1253--1306
  \urlprefix\url{http://link.aps.org/doi/10.1103/RevModPhys.84.1253}

\bibitem{2011_Kozlowski_JSTAT_P03018}
Kozlowski K~K, Maillet J~M and Slavnov N~A 2011 {\em Journal of Statistical
  Mechanics: Theory and Experiment\/} {\bf 2011} P03018
  \urlprefix\url{http://stacks.iop.org/1742-5468/2011/i=03/a=P03018}

\bibitem{2008_Rigol_NATURE_452}
Rigol M, Dunjko V and Olshanii M 2008 {\em Nature\/} {\bf 452} 854--858

\bibitem{2007_Rigol_PRL_98}
Rigol M, Dunjko V, Yurovsky V and Olshanii M 2007 {\em Phys. Rev. Lett.\/} {\bf
  98}(5) 050405
  \urlprefix\url{http://link.aps.org/doi/10.1103/PhysRevLett.98.050405}

\bibitem{2015_Ilievski_PRL_115}
Ilievski E, De~Nardis J, Wouters B, Caux J~S, Essler F~H~L and Prosen T 2015
  {\em Phys. Rev. Lett.\/} {\bf 115}(15) 157201
  \urlprefix\url{http://link.aps.org/doi/10.1103/PhysRevLett.115.157201}

\bibitem{2015_Essler_PRA_91}
Essler F~H~L, Mussardo G and Panfil M 2015 {\em Phys. Rev. A\/} {\bf 91}(5)
  051602 \urlprefix\url{http://link.aps.org/doi/10.1103/PhysRevA.91.051602}

\bibitem{2013_Fagotti_PRB_87}
Fagotti M and Essler F~H~L 2013 {\em Phys. Rev. B\/} {\bf 87}(24) 245107
  \urlprefix\url{http://link.aps.org/doi/10.1103/PhysRevB.87.245107}

\bibitem{2012_Caux_PRL_109}
Caux J~S and Konik R~M 2012 {\em Phys. Rev. Lett.\/} {\bf 109}(17) 175301
  \urlprefix\url{http://link.aps.org/doi/10.1103/PhysRevLett.109.175301}

\bibitem{2012_Essler_PRL_109}
Essler F~H~L, Evangelisti S and Fagotti M 2012 {\em Phys. Rev. Lett.\/} {\bf
  109}(24) 247206
  \urlprefix\url{http://link.aps.org/doi/10.1103/PhysRevLett.109.247206}

\bibitem{2012_Mossel_JPA_45}
Mossel J and Caux J~S 2012 {\em Journal of Physics A: Mathematical and
  Theoretical\/} {\bf 45} 255001
  \urlprefix\url{http://stacks.iop.org/1751-8121/45/i=25/a=255001}

\bibitem{2011_Cassidy_PRL_106}
Cassidy A~C, Clark C~W and Rigol M 2011 {\em Phys. Rev. Lett.\/} {\bf 106}(14)
  140405 \urlprefix\url{http://link.aps.org/doi/10.1103/PhysRevLett.106.140405}

\bibitem{2013_Caux_PRL_110}
Caux J~S and Essler F~H~L 2013 {\em Phys. Rev. Lett.\/} {\bf 110}(25) 257203
  \urlprefix\url{http://link.aps.org/doi/10.1103/PhysRevLett.110.257203}

\bibitem{2013_Mussardo_PRL_111}
Mussardo G 2013 {\em Phys. Rev. Lett.\/} {\bf 111}(10) 100401
  \urlprefix\url{http://link.aps.org/doi/10.1103/PhysRevLett.111.100401}

\bibitem{2013_Pozsgay_JSTAT_P07003}
Pozsgay B 2013 {\em Journal of Statistical Mechanics: Theory and Experiment\/}
  {\bf 2013} P07003
  \urlprefix\url{http://stacks.iop.org/1742-5468/2013/i=07/a=P07003}

\bibitem{2016_Alba_JSTAT_P043105}
Alba V and Calabrese P 2016 {\em Journal of Statistical Mechanics: Theory and
  Experiment\/} {\bf 2016} 043105
  \urlprefix\url{http://stacks.iop.org/1742-5468/2016/i=4/a=043105}

\bibitem{2013_Kormos_PRB_88}
Kormos M, Shashi A, Chou Y~Z, Caux J~S and Imambekov A 2013 {\em Phys. Rev.
  B\/} {\bf 88}(20) 205131
  \urlprefix\url{http://link.aps.org/doi/10.1103/PhysRevB.88.205131}

\bibitem{2014_Fagotti_PRB_89}
Fagotti M, Collura M, Essler F~H~L and Calabrese P 2014 {\em Phys. Rev. B\/}
  {\bf 89}(12) 125101
  \urlprefix\url{http://link.aps.org/doi/10.1103/PhysRevB.89.125101}

\bibitem{2014_Wouters_PRL_113}
Wouters B, De~Nardis J, Brockmann M, Fioretto D, Rigol M and Caux J~S 2014 {\em
  Phys. Rev. Lett.\/} {\bf 113}(11) 117202
  \urlprefix\url{http://link.aps.org/doi/10.1103/PhysRevLett.113.117202}

\bibitem{2014_Pozsgay_PRL_113}
Pozsgay B, Mesty\'an M, Werner M~A, Kormos M, Zar\'and G and Tak\'acs G 2014
  {\em Phys. Rev. Lett.\/} {\bf 113}(11) 117203
  \urlprefix\url{http://link.aps.org/doi/10.1103/PhysRevLett.113.117203}

\bibitem{2014_Kormos_PRA_89}
Kormos M, Collura M and Calabrese P 2014 {\em Phys. Rev. A\/} {\bf 89}(1)
  013609 \urlprefix\url{http://link.aps.org/doi/10.1103/PhysRevA.89.013609}

\bibitem{2014_DeNardis_PRA_89}
De~Nardis J, Wouters B, Brockmann M and Caux J~S 2014 {\em Phys. Rev. A\/} {\bf
  89}(3) 033601
  \urlprefix\url{http://link.aps.org/doi/10.1103/PhysRevA.89.033601}

\bibitem{2014_Sotiriadis_JSTAT_P07024}
Sotiriadis S and Calabrese P 2014 {\em Journal of Statistical Mechanics: Theory
  and Experiment\/} {\bf 2014} P07024
  \urlprefix\url{http://stacks.iop.org/1742-5468/2014/i=7/a=P07024}

\bibitem{2015_Mestyan_JSTAT_P0400}
Mestyán M and Pozsgay B 2014 {\em Journal of Statistical Mechanics: Theory and
  Experiment\/} {\bf 2014} P09020
  \urlprefix\url{http://stacks.iop.org/1742-5468/2014/i=9/a=P09020}

\bibitem{2016_Caux_arxiv_1603.04689}
{Caux} J~S 2016 {\em ArXiv e-prints\/} (\textit{Preprint} \eprint{1603.04689})

\bibitem{2013_Eriksson_JPA_46}
Eriksson E and Korepin V 2013 {\em Journal of Physics A: Mathematical and
  Theoretical\/} {\bf 46} 235002
  \urlprefix\url{http://stacks.iop.org/1751-8121/46/i=23/a=235002}

\bibitem{2014_Fokkema_PRA_89}
Fokkema T, Eli\"ens I~S and Caux J~S 2014 {\em Phys. Rev. A\/} {\bf 89}(3)
  033637 \urlprefix\url{http://link.aps.org/doi/10.1103/PhysRevA.89.033637}

\bibitem{2016_Vlijm_SCIPOST_inprep}
{R}ogier {V}lijm, {I}an~{S}ebastiaan {E}li\"ens and {J}ean-{S}\'ebastien {C}aux
  {\em arXiv:1606.09516\/} \urlprefix\url{http://arxiv.org/abs/1606.09516}

\bibitem{1963_Lieb_PR_130}
Lieb E~H and Liniger W 1963 {\em Phys. Rev.\/} {\bf 130}(4) 1605--1616
  \urlprefix\url{http://link.aps.org/doi/10.1103/PhysRev.130.1605}

\bibitem{1998_Cheon_PLA_243}
Cheon T and Shigehara T 1998 {\em Physics Letters A\/} {\bf 243} 111 -- 116
  ISSN 0375-9601
  \urlprefix\url{http://www.sciencedirect.com/science/article/pii/S0375960198001881}

\bibitem{1999_Cheon_PRL_82}
Cheon T and Shigehara T 1999 {\em Phys. Rev. Lett.\/} {\bf 82}(12) 2536--2539
  \urlprefix\url{http://link.aps.org/doi/10.1103/PhysRevLett.82.2536}

\bibitem{1989_Izergin_JPA_22}
Izergin A~G, Korepin V~E and Reshetikhin N~Y 1989 {\em Journal of Physics A:
  Mathematical and General\/} {\bf 22} 2615
  \urlprefix\url{http://stacks.iop.org/0305-4470/22/i=13/a=052}

\bibitem{1990_Tsvelik_PRB_42}
Tsvelik A~M 1990 {\em Phys. Rev. B\/} {\bf 42}(1) 779--785
  \urlprefix\url{http://link.aps.org/doi/10.1103/PhysRevB.42.779}

\bibitem{1992_Frahm_JPA_25}
Frahm H 1992 {\em Journal of Physics A: Mathematical and General\/} {\bf 25}
  1417 \urlprefix\url{http://stacks.iop.org/0305-4470/25/i=6/a=005}

\bibitem{1987_Pokrovsky_ZETF_93}
Pokrovsky S and Tsvelik A 1987 {\em Zh. Eksp.Theor. Fiz.\/} {\bf 93} 2232--2246

\bibitem{1997_Frahm_JPA_30}
Frahm H and Rödenbeck C 1997 {\em Journal of Physics A: Mathematical and
  General\/} {\bf 30} 4467
  \urlprefix\url{http://stacks.iop.org/0305-4470/30/i=13/a=005}

\bibitem{1990_Frahm_JPA_23}
Frahm H and Yu N~C 1990 {\em Journal of Physics A: Mathematical and General\/}
  {\bf 23} 2115 \urlprefix\url{http://stacks.iop.org/0305-4470/23/i=11/a=032}

\bibitem{2000_Zvyagin_LTP_26}
Zvyagin A~A 2000 {\em Low Temperature Physics\/} {\bf 26} 134--146
  \urlprefix\url{http://scitation.aip.org/content/aip/journal/ltp/26/2/10.1063/1.593878}

\bibitem{2001_Zvyagin_EPJB_19}
Zvyagin A, Kl{\"u}mper A and Zittartz J 2001 {\em The European Physical Journal
  B-Condensed Matter and Complex Systems\/} {\bf 19} 25--36

\bibitem{2003_Zvyagin_PRB_68}
Zvyagin A~A and Kl\"umper A 2003 {\em Phys. Rev. B\/} {\bf 68}(14) 144426
  \urlprefix\url{http://link.aps.org/doi/10.1103/PhysRevB.68.144426}

\bibitem{1990_Frahm_PRB_42}
Frahm H and Korepin V~E 1990 {\em Phys. Rev. B\/} {\bf 42}(16) 10553--10565
  \urlprefix\url{http://link.aps.org/doi/10.1103/PhysRevB.42.10553}

\bibitem{HubbardBOOK}
Essler F~H~L, Frahm H, G\"ohmann F, Kl\"umper A and Korepin V~E 2005 {\em {The
  One-Dimensional Hubbard Model}\/} (Cambridge University Press)

\end{thebibliography}

\end{document}